\documentclass[12pt]{iopart}

\expandafter\let\csname equation*\endcsname\relax
\expandafter\let\csname endequation*\endcsname\relax

\usepackage{graphicx}
\usepackage{cite}
\usepackage{amsmath,amssymb,amsfonts}
\usepackage{multirow}
\usepackage{makecell}
\usepackage{tabu}
\usepackage{caption}
\usepackage{subcaption}
\usepackage[colorlinks]{hyperref}
\begin{document}

\title[]{Category Guided Attention Network for Brain Tumor Segmentation in MRI}

\author{
    Jiangyun Li\textsuperscript{1,{*}},
	Hong Yu\textsuperscript{1},
	Chen Chen\textsuperscript{2},
	Meng Ding\textsuperscript{3},
	Sen Zha\textsuperscript{1}
}

\address{
$^1$ School of Automation and Electrical Engineering, University of Science and Technology Beijing, Beijing 100083, China
}

\address{
$^2$ Center for Research in Computer Vision, University of Central Florida, Orlando, USA
}

\address{
$^3$ Scoop Medical, Houston, USA
}

\eads{\mailto{leejy@ustb.edu.cn}, \mailto{g20198754@xs.ustb.edu.cn}, \mailto{chen.chen@ucf.edu}, \mailto{meng.ding@okstate.edu}, \mailto{g20198675@xs.ustb.edu.cn},}

\vspace{10pt}

\begin{indented}
\item[]September 2021
\end{indented}

\begin{abstract}

\textbf{Objective}: Magnetic resonance imaging (MRI) has been widely used for the analysis and diagnosis of brain diseases. Accurate and automatic brain tumor segmentation is of paramount importance for radiation treatment. However, low tissue contrast in tumor regions makes it a challenging task.

\textbf{Approach}: We propose a novel segmentation network named Category Guided Attention U-Net (CGA U-Net). In this model, we design a Supervised Attention Module (SAM) based on the attention mechanism, which can capture more accurate and stable long-range dependency in feature maps without introducing much computational cost. Moreover, we propose an intra-class update approach to reconstruct feature maps by aggregating pixels of the same category.

\textbf{Main results}: Experimental results on the BraTS 2019 datasets show that the proposed method outperformers the state-of-the-art algorithms in both segmentation performance and computational complexity.

\textbf{Significance}: The CGA U-Net can effectively capture the global semantic information in the MRI image by using the SAM module, while significantly reducing the computational cost. Code is available at \url{https://github.com/delugewalker/CGA-U-Net}.
\end{abstract}

\vspace{2pc}
\noindent{\it Keywords}: Segmentation, Attention, Brain Tumor, MRI, Intra-Class, Inter-Class.
%\maketitle

\section{Introduction}

Gliomas are the most common primary brain malignancies and are a serious threat to public health. Accurate identification of tumors can help doctors achieve more reasonable diagnosis and treatment strategies. As a high-accuracy identification method, tumor segmentation can help in a variety of clinical applications, such as radiation therapy planning \cite{kiljunen2020deep, kuisma2020validation, agn2019modality}, tumor growth monitoring \cite{kickingereder2019automated}, intra-operative assistance \cite{hollon2020near, tanzi2021real}, etc. However, it is very time-consuming to manually label the tumor location by medical experts, and the labeling accuracy may be insufficient. Therefore, it is very important to develop an automatic segmentation method. However, the contour of the lesion area can only be determined by the change of the pixel intensity relative to the surrounding tissues. In addition, many MR images have gray offset fields such as motion artifact which may cause either ghost images or diffuse image noise, making it difficult to accurately segment the contour. The variation in size, extension, and position of tumor also makes segmentation algorithms unable to use strong prior constraints similar to normal tissue in shape and location parameters.

Recently, convolutional neural networks (CNN) have made significant progress in solving challenging tasks such as classification, segmentation and object detection \cite{he2016deep, girshick2014rich, long2015fully}. Fully convolutional network (FCN) \cite{long2015fully} first achieves end-to-end semantic segmentation with satisfactory results. It also has an excellent performance in medical image segmentation. Ronneberger et al. \cite{ronneberger2015u} propose U-Net with the encoder-decoder structure to generate a dense prediction of the entire image. After that, a variety of U-Net variants based on the encoder-decoder architecture have emerged in an endless stream, and have achieved convincing results \cite{xiao2018weighted, cai2020dense, ibtehaz2020multiresunet, zhou2018unet++, huang2020unet}. For brain tumor segmentation task, the state-of-the-art methods \cite{nuechterlein20183d, chen2018s3d, isensee2018no, myronenko20183d, jiang2019two, islam2019brain, chen20193d, xu2019deep} are based on the 3D U-Net \cite{cciccek20163d} which is a 3D encoder-decoder architecture with skip connections.

However, it is difficult to establish connections between long-distance voxels due to the small size of convolution kernel. Same as long-distance dependence, we attempt to capture global semantic information in medical images, which refers to the relationship between any two points in the image/matrix with no consideration of the physical distance between them. However, this limitation of convolution kernel brings difficulties to learn global semantic information which can effectively guide the model to segment. Moreover, in the process of establishing connections between distant pixels, over-segmentation or under-segmentation will be suppressed. In order to capture long-range dependency in feature map, one possible way is to repeatedly stack convolutional layers to increase the receptive field. However, this solution is computationally expensive, especially in 3D convolution and may result in model optimization difficulties. Yu et al. \cite{yu2015multi} try to solve this problem using dilated convolution, but still only expand the receptive field at the local feature scale. Stacked dilated convolution will also cause gridding effect and loss the continuity of information. This is fatal for the task of pixel-level dense prediction such as semantic segmentation.

The attention mechanism \cite{vaswani2017attention} has been increasingly explored to model long-range dependency for computer vision tasks \cite{wang2018non, zhu2019asymmetric, fu2019dual, huang2019ccnet, yu2020context}. The essence of it is to get the weights that describe the similarity between features, and then re-weight the features by these weights. For instance, in semantic segmentation, Non-local neural networks \cite{wang2018non} perform two matrix multiplications, one for generating attention map and the other for updating original feature maps, which is a typical self-attention module (Figure.~\ref{fig:comparison-attention.sub1}). The attention map calculated by the first matrix multiplication represents the similarity (relationship) between each pixel in the feature map and any other pixel. In fact, this similarity matrix can be seen as a kind of auto-correlation, targeting to find the parts with high similarity in itself. In the second step, the matrix multiplication uses this similarity as a weighting coefficient to weight the original feature map to obtain a new feature map enhanced by global information. The attention mechanism is also used in medical image segmentation and previous works have proved the importance of long distance dependency \cite{oktay2018attention, zhou2020one, zhou2020multi}. Oktay et al.  \cite{oktay2018attention} combined the self-attention mechanism with 3D-UNet, and successfully guided the network to model the global information through the attention gating module embedded in skip connections. However, none of these methods can escape the matrix multiplication framework, which has two problems -- instability of the attention map and high computational complexity. Tan et al. \cite{tan2020explicitly} describes the problem that the attention map is not intuitive. Their visualization results of the attention map show that content-dependent key and query play a minor role in the final attention-maps. In addition, Zheng et al. \cite{zheng2019looking} proved that it is difficult to learn multiple consistent (i.e., attending on the same part for each sample) attention maps without annotations in the attention mechanism. High computational complexity is a recognized problem for attention mechanism. CCNet \cite{huang2019ccnet} explains that for dense prediction tasks such as semantic segmentation, where high-resolution feature maps are required. However, non-local (matrix multiplication) based attention methods have a high computational complexity of $O(N^2)$ for 2D images and consume a large GPU memory footprint. This problem is more severe on 3D images.

CPNet \cite{yu2020context} introduces a context prior (CP) layer based on the attention mechanism, which contains a CP map similar to attention map. Due to the lack of explicit modeling, the attention map is difficult to generate clear structural information. Therefore, the CP map is obtained via a convolution layer in a supervised manner instead of matrix multiplication, making this similarity matrix more accurate and stable. However, the CP map comes from a local feature map with only one convolution directly, which means parameters before the feature map are shared for optimizing the CP map and label prediction and will cause learning difficulties during the training process. Moreover, the size of the CP map is still large and the update method of the original feature map is remains as matrix multiplication. 

Although the aforementioned attention mechanisms can improve model’s capability of capturing long-range dependency, there are some challenges in the implementation process. In order to overcome these limitations, our work can be summarized as follows:

% \begin{itemize}[noitemsep,leftmargin=*]
1) We propose a novel Supervised Attention Module (SAM) with an attention convolution (conv) path to construct the attention map from input images by using a series of convolutions and utilizing the accurate guidance from ground truth supervision.

2) To reduce the computational complexity, we redefine the form of attention map in SAM to have a much smaller size. Moreover, we design a new approach called “intra-class update” to reconstruct the feature map according to the category without using matrix multiplication. These two improvements significantly reduce the computational cost.

3) The general attention mechanism aggregates intra-class information through the similarity between pixels but ignores the differences between classes. To this end, we propose a novel inter-class distance optimization method to expand distances among classes, making it easier to distinguish pixels of different classes.
% \end{itemize}

\begin{figure}[ht]
 % Caption and label go in the first argument and the figure contents
 % go in the second argument
    \centering
    % \vspace{-0.5cm}
    % \setlength\belowdisplayskip{3pt}
    \includegraphics[width=1\linewidth]{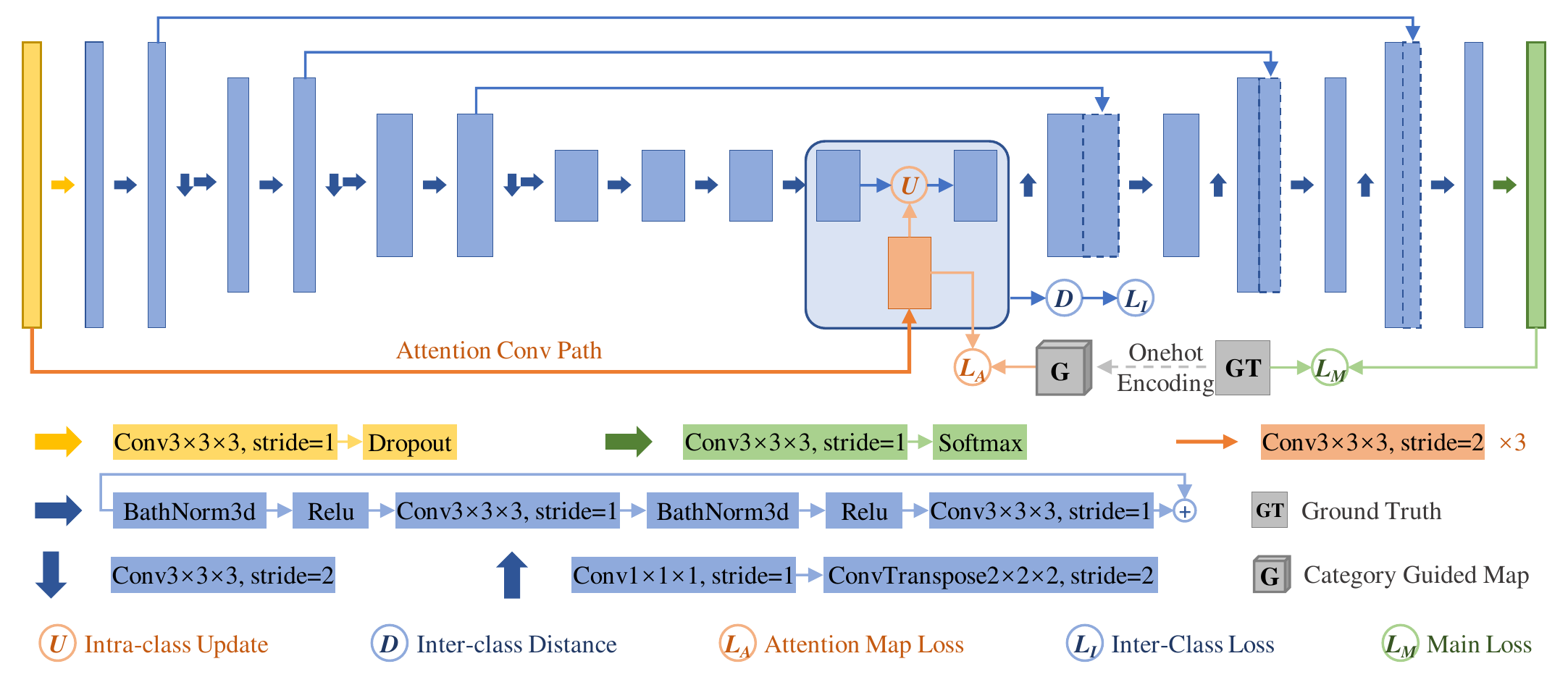}
    \caption{Overview of the proposed category guided attention U-Net (CGA U-Net).}
    \label{fig:overview}
\end{figure}

Our proposed segmentation approach is called Category Guided Attention U-Net (CGA U-Net) (Figure.~\ref{fig:overview}) and achieves the state-of-the-art performance on the BraTS2019 dataset. Next, we articulate CGA U-Net in the following section. 

\section{Related work}

\subsection{Medical Image Segmentation}
Medical image segmentation aims to automatically divide various areas in medical images to help doctors achieve more efficient and accurate diagnoses. Early image segmentation methods including thresholding \cite{manikandan2014multilevel}, clustering-based \cite{gomez2015fuzzy}, etc. have achieved some effects in specific circumstances. However, different from RGB images, medical images have the characteristics of a single scene, low contrast and more noise. Therefore, these methods cannot satisfactorily solve the problem of medical image segmentation. In recent years, algorithms based on convolutional neural networks have been explored in the field of medical image segmentation and have achieved good results \cite{ronneberger2015u, xiao2018weighted, cai2020dense, ibtehaz2020multiresunet, zhou2018unet++, huang2020unet, cciccek20163d, milletari2016v, nuechterlein20183d, chen2018s3d, isensee2018no, myronenko20183d, jiang2019two, islam2019brain, chen20193d, xu2019deep}. 

For both 2D and 3D medical image segmentation tasks, most methods adopted encoder-decoder architecture with skip connections based on U-Net \cite{ronneberger2015u}. The decoder can gradually upsample the features extracted by the encoder to the original size, thereby achieving pixel-level classification. To recover the pixel-level classification map of the image, \cite{xiao2018weighted, cai2020dense, ibtehaz2020multiresunet} try to modify the basic unit and upsampling structure in encoder-decoder. \cite{zhou2018unet++, huang2020unet} enable the model to extract richer feature representations by improving the skip connection method.

\subsection{Attention Mechanism}
The attention mechanism can be seen as a technique for incorporating global information by finding the relationship between any two pixels in an image. This long-range dependency is critical for pixel-level tasks like semantic segmentation.
Non-local \cite{wang2018non} developed a template of two-step matrix multiplications for visual attention and was widely adopted by subsequent methods. 
Asymmetric non-local method \cite{zhu2019asymmetric} samples the pixels of feature maps, thereby reducing the computation and memory cost without sacrificing the performance. DANet \cite{fu2019dual} combines spatial attention with channel attention to capture long-range dependency among feature maps. CCNet \cite{huang2019ccnet} reduces the computation by decomposing the two dimensions of the attention map while retaining good results.

\begin{figure}[ht]
    \centering
    
    \subcaptionbox{Self Attention Module \label{fig:comparison-attention.sub1}}{
        \includegraphics[width=0.7\textwidth]{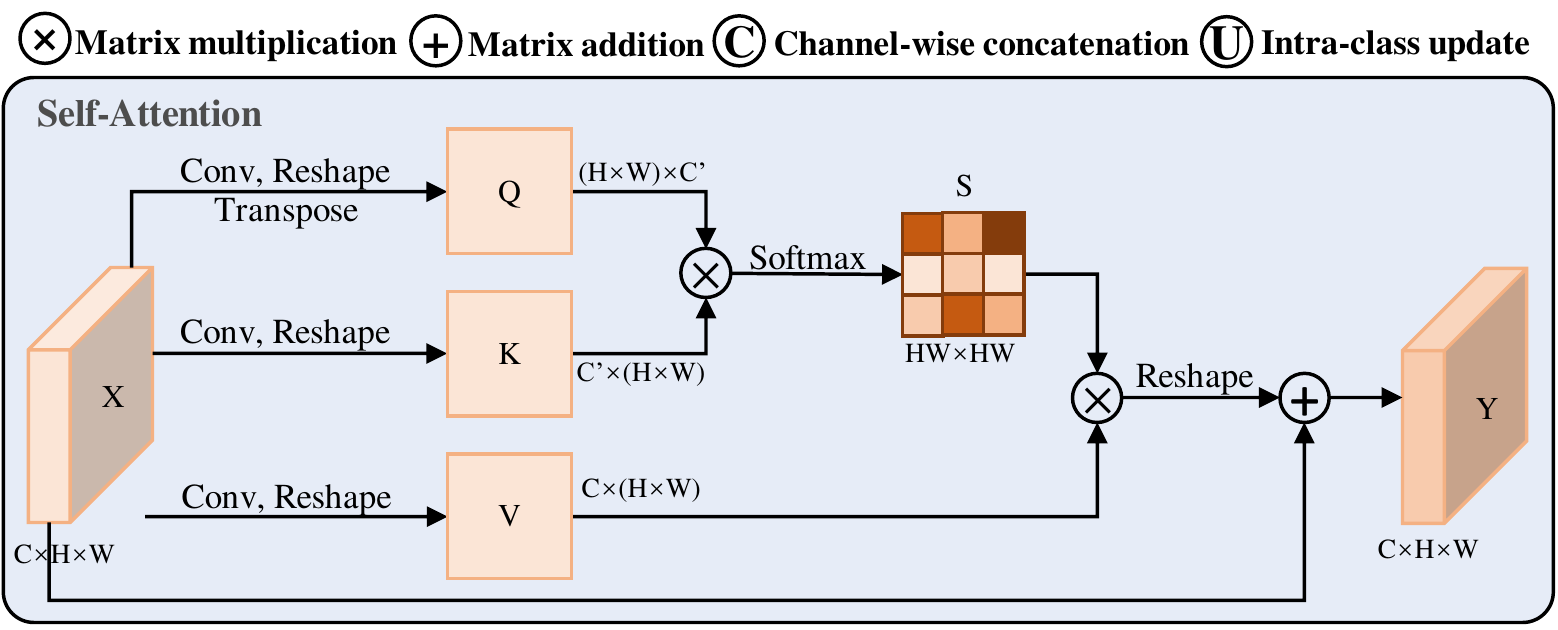}
    }
    % \hfill
    \subcaptionbox{CP Layer \label{fig:comparison-attention.sub2}}{
        \includegraphics[width=0.7\textwidth]{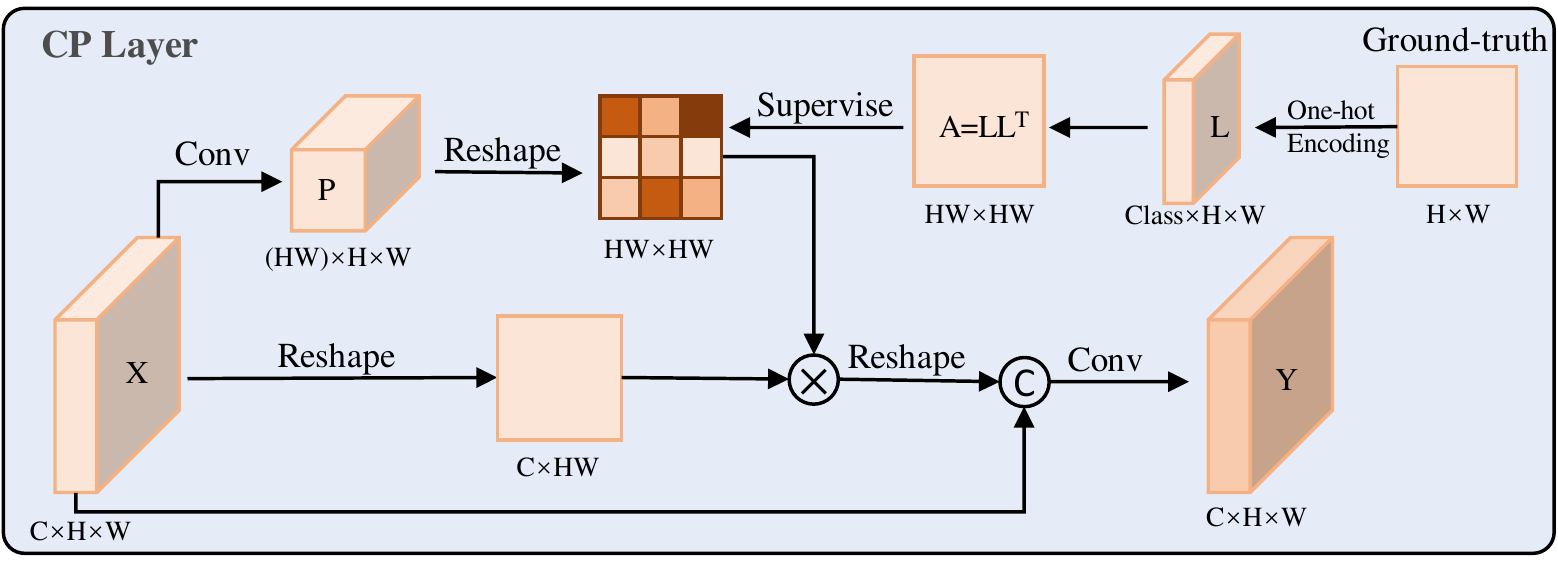}
    }
    % \hfill
    \subcaptionbox{Supervised Attention Module \label{fig:comparison-attention.sub3}}{
        \includegraphics[width=0.7\textwidth]{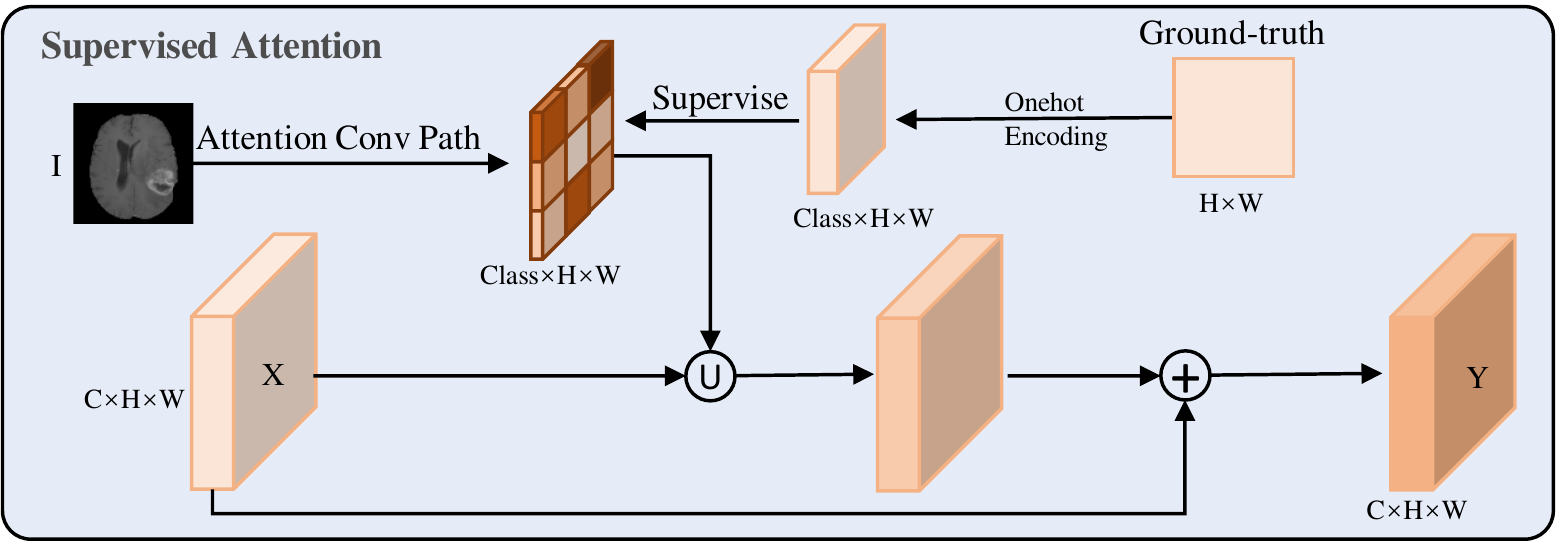}
    }
    
    \caption{Comparison of different attention modules. (a) classic self-attention module, (b) context prior layer \cite{yu2020context}, and (c) the proposed supervised attention module (SAM).}
    \label{fig:comparison-attention}
\end{figure}

A typical self-attention structure is shown in Figure.~\ref{fig:comparison-attention.sub1}. Without loss of generality, the module is described in 2D here for notation clarity. Extension to 3D is straightforward. Consider a local feature map $X\in\mathbb{R}^{C\times H \times W}$, where $C$, $H$ and $W$ indicate the channel number, spatial height and width, respectively. The attention module first applies three 1$\times$1 convolutions {$W_{q}$}, {$W_{k}$}, {$W_{v}$} and a reshape operation {$R$} to generate three matrices $\{Q, K, V\}\in\mathbb{R}^{C^{'}\times N}$ which can be written as 

\begin{equation}
\small
Q=R(W_{q}(X)),  K=R(W_{k}(X)),  V=R(W_{v}(X)),
\end{equation}
where $C^{'}$ is the channel number after the convolution and $N=H\times W$. Then a matrix multiplication and the softmax normalization function {$f$} are applied to $Q$ and $K^T$ to obtain the attention map $S\in\mathbb{R}^{N \times N}$ as

\begin{equation}
\small
S=f(Q \times K^T).
\end{equation}
With this similarity matrix, we can reconstruct matrix $V$ using another multiplication and reshape operation. Finally, the output $Y\in\mathbb{R}^{C \times H \times W}$ is given by 

\begin{equation}
\small
Y=R(V \times S) + X.
\end{equation}
This attention structure has some inherent limitations. On one hand, it is computationally expensive given the large size of attention map. On the other hand, experiments implemented by CPNet \cite{yu2020context} prove that the attention map lacking an explicit modeling cannot produce effective structural information for object segmentation, which has a shortcoming of instability and may result in inaccurate description.

Instead of matrix multiplication, context prior (CP) layer \cite{yu2020context} employs convolution and reshape operations to get the CP map $P$ with the same dimension as $S$, as illustrated in Figure.~\ref{fig:comparison-attention.sub2}. The process can be formulated as

\begin{equation}
\small
P = R(Conv(X)).
\end{equation}
At the same time, the down-sampled ground truth $GT$ is encoded into $L \in \mathbb{R}^{C_l \times H \times W}$ with the one-hot encoding, where $C_l$ denotes the number of classes. Then the affinity map $A \in \mathbb{R}^{C_l \times H \times W}$ is produced by a multiplication between $L$ and $L^T$, i.e.

\begin{equation}
\small
L = Onehot(GT), \quad A = L \times L^T.
\end{equation}
The CP layer uses the affinity map to supervise the CP map, which can correct the unstable description and improve the accuracy. However, the scale of the CP map is still quite large, and there will be a lot of redundant information when updating the local feature map using multiplication. The CP map actually attempt to establish a fully connected relationship between all pixels and the pairwise similarity measurement leads to massive consumption of computation and memory cost. In fact, describing the information among pixels of the same category does not require such a large-scale fully connected graph of pairwise relationship. There are some works \cite{huang2019ccnet, chen2020compressed} aiming to eliminate redundant information and reduce consumption of computation by compressing the attention calculation process, which is effective to a certain extent. 

Moreover, the CP layer converts the feature map $X\in\mathbb{R}^{C \times H \times W}$ into $P\in\mathbb{R}^{HW\times H \times W}$ through a convolution layer, which will cause the physical meaning of the feature map to change. The CP map (size of $HW \times HW$) describes the relationship between any two points in the image (a total of $H \times W$ points), but the first dimension of CP map is actually the channel dimension, which does not well represent the relationship between a certain point and other points (spatial dimensions). This may give rise to difficulties in optimizing parameters. Getting CP map from local feature map directly through a convolution will also cause conflicts between the affinity loss used for supervising CP map and the loss for class label prediction in back propagation because they share the parameters before local feature map.

For medical image segmentation, there are also some methods that attempt to combine the attention mechanism with the encoder-decoder structure \cite{oktay2018attention, abraham2019novel, jiang2019ahcnet, ni2019raunet, sinha2020multi}. Attention U-Net \cite{oktay2018attention} uses the attention gate module to pay attention to target structures of varying shapes and sizes. Multi-scale self-guided attention \cite{sinha2020multi} eliminates redundant information and focuses on discriminative features by modeling richer contextual dependencies with multi-scale self-attention. 

The major step in the attention mechanism is to obtain the attention weight. However, due to the lack of supervision information, these aforementioned attention algorithms based on the matrix multiplication framework may cause the instablity of the attention weight. Moreover, the calculational burden of matrix multiplication also makes the implementation blocked.

\subsection{Intra-class \& inter-class methods}

In computer vision classification tasks, the minimization of intra-class distance is used to ensure  consistency within a certain class, and the optimization method of inter-class distance is used to amplify the difference between classes. Ming et al. \cite{ming2017simple} design a class-wise triplet loss based on intra/inter-class distance metric learning. By calculating the center of the positive/negative sample, the intra-class distance of the feature is minimized and the inter-class distance is maximized. 

Semantic segmentation is a pixel-level classification task. It is necessary to enhance the intra-class consistency between pixels and optimize the inter-class distance to obtain more robust segmentation result. Yu et al. \cite{yu2018learning} use multi-scale information and global average pooling to implicitly solve the problem of inconsistencies within the class. Moreover, the boundary information is adopted to enhance inter-class distinction.

Different from the above methods, we use global semantic information to explicitly find pixels belonging to the same category with the help of a supervised attention map. By finding the prototype vector of each category, on the one hand, the auto-correlation of the attention mechanism is realized through re-weighting, and on the other hand, the distance between classes is optimized.

\section{Methodology}
% ======================================================================

%FIXME
% The next section should start with the heading “Methods” (not “Materials and Methods”). It should succinctly describe the techniques and instrumentation used, define the patient population, etc.
% \begin{equation}
% \mathrm{Test for equation numbering - has to be with brackets}
% \end{equation}

\subsection{Overview of the Category Guided Attention U-Net}
An overview of the proposed Category Guided Attention U-Net (CGA U-Net) is presented in Figure.~\ref{fig:overview}. The baseline we used in experiments (including ablation study) is the U-Net with some modifications, and the details of our network structure are provided in Table.~\ref{tab:network_details}. Specifically, an attention convolution path is injected to generate the attention map from the input. Then the attention map is utilized to guide the intra-class update process of the feature map and optimize the inter-class distance simultaneously.

There are three losses in our network. The main loss uses softmax dice loss to optimize the segmentation results and the attention map loss adopts mean squared error (MSE) loss for supervising the attention map. As mentioned before, we design $loss_{inter}$ to optimize the inter-class distance. These three losses are denoted as $L_{M}$, $L_{A}$, $L_{I}$ respectively in Figure.~\ref{fig:overview}. In the training process, the optimization of loss is divided into two stages. In the first 20 epochs, we add $L_{M}$ and $L_{A}$ together as the overall loss while after 20 epochs the overall loss is the sum of the three losses. We set a weight for $L_{I}$ to keep it at the same order of magnitude as the other two loss functions. In addition, the weights of $L_{M}$ and $L_{A}$ are the same and we fixed them in all experiments.

To avoid optimizing difficulties caused by the conversion aforementioned, we redefine the form of attention map to have a smaller size and channel dimension. An independent attention convolution (conv) path is added to eliminate conflicts between the two losses. With the new form of attention map, we design new methods to update the feature map based on categories, which can reduce the computational complexity.

% \subsection{Supervised Attention Module}
\subsection{New Attention Map Form}
% \textbf{New Attention Map Form.}
In the proposed SAM, we develop an attention path to directly obtain the attention map from the original input $I$ through a series of convolutions, as demonstrated in Figure.~\ref{fig:comparison-attention.sub3}. Different from self-attention and context prior (CP) block, the size of our attention map is decreased from $N \times N$ to $C_{l} \times H \times W$ ($C_l$: the number of classes) and the attention map is represented as 

\begin{equation}
\small
S = Conv(I).
\end{equation}

Accordingly, we encode the down-sampled ground truth into one-hot form to get the category guided map $G$, which has the same size as $S$. Each $1 \times C_{l}$ vector in $G$ is composed of a single high value (1) while the others are low (0). Each channel of $G$ represents the pixel distribution of a single category. Then the attention map is supervised by the category guided map for learning category-based attention information. The attention map arranged by category can more accurately aggregate pixels of the same class and differentiate pixels of different classes. 

\begin{figure}[ht]
    \centering
    
    \subcaptionbox{Intra-class update method \label{fig:intra.sub1}}{
        \includegraphics[width=0.45\textwidth]{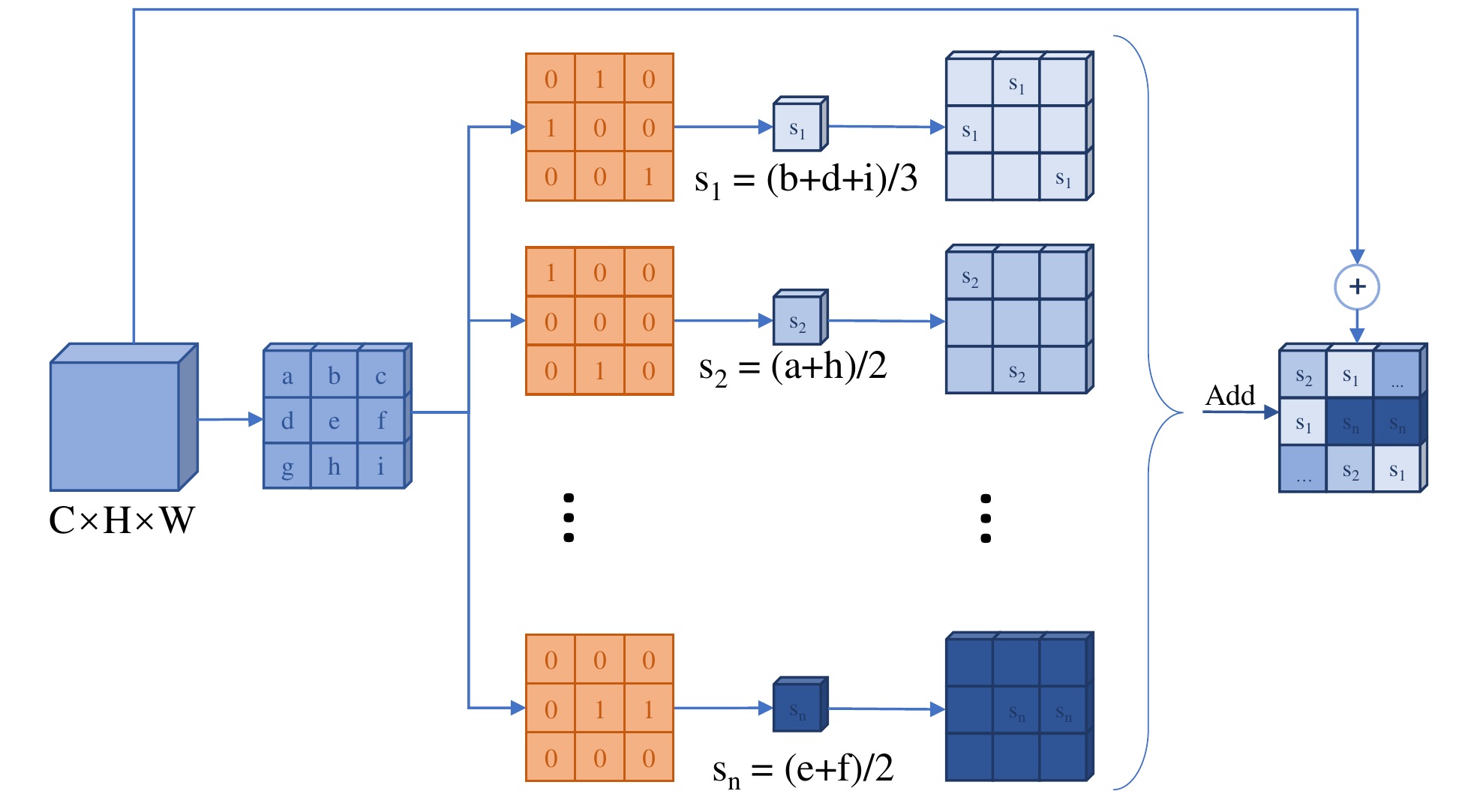}
    }
    \hfill
    \subcaptionbox{Inter-class distance optimization \label{fig:intra.sub2}}{
        \includegraphics[width=0.45\textwidth]{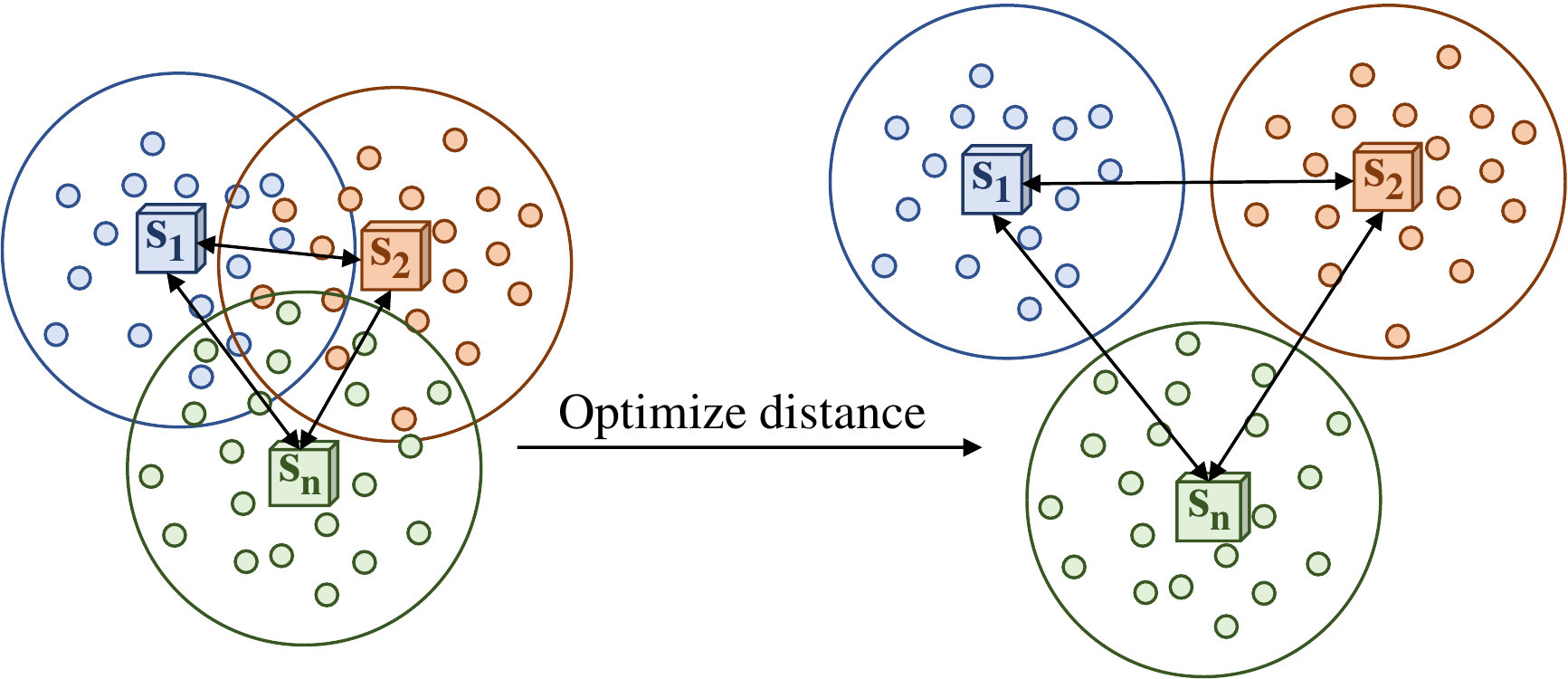}
    }

    \caption{Intra-class update method and Inter-class distance optimization.}
    \label{fig:intra}
\end{figure}

\subsection{Intra-class Update Method}
% \textbf{Intra-class Update Method.}
After obtaining the attention map, we propose a novel category-based approach to update the local feature map called the intra-class update method. As illustrated in Figure.~\ref{fig:intra.sub1}, given a local feature map $X\in\mathbb{R}^{C\times H \times W}$, we regard each pixel as a $1 \times C$ vector, and there are a total of $H \times W$ vectors. Then the attention map is disassembled according to categories, and each category corresponds to a mask $M_k\in\mathbb{R}^{H \times W}$. For each mask, we generate the corresponding prototype vector $p_k$ through masked average pooling. Specifically, we take out the feature vectors of pixels belonging to a certain category, and calculate their average value as the prototype vector of this category:

\begin{equation}
\small
p_k = \frac{1}{\sum_{i=1}^{H \times W} w_i} \sum_{i=1}^{H \times W}x_i \times w_i,
\end{equation}
where $x_i$ is the vector in feature map, $w_i$ is the weight in mask and $i$ indicates the position. Then the prototype vector of each category is mapped back to the size of the local feature map according to the mask as

\begin{equation}
y_i = map(p_k, M_k),
\end{equation}
where $y_i$ is the vector in the feature map after mapping. Finally, the regenerated feature maps of all the categories are added to obtain the final result. Through this intra-class update method, we can effectively aggregate pixels of the same category.

\subsection{Inter-class Distance Optimization}
% \textbf{Inter-class Distance Optimization.}
Considering that there may be similar pixels among different categories, we propose an inter-class distance optimization method (Figure.~\ref{fig:intra.sub2}) to make it easier for distinguishing these pixels.

Based on the existing prototype vector of each category, we compute the Euclidean distance between every two of them. The sum of all distances represents the inter-class distance of the feature map and is denoted as

\begin{equation}
\small
D = \sum_{0<i, j\le Cl} E(p_i, p_j), 
\end{equation}
where $E$ denotes the Euclidean distance between two vectors. Our optimization goal is to maximize the inter-class distance, so we define an inter-class loss as follows

\begin{equation}
loss_{inter} = \log \frac{1}{1+D},
\end{equation}
where $D$ represents the inter-class distance. We can minimize $loss_{inter}$ in the training process in order to optimize the distance.

\begin{table*}[h]
\scriptsize
    \centering
    \caption{Detailed description of our network structure.}
    \label{tab:network_details}
    \setlength{\tabcolsep}{3mm}{
    \begin{tabular}{c|c|c|c|c}
        \hline
        % \multirow{2}{*}{Method} & \multicolumn{3}{c}{Dice Score (\%) $\uparrow$} & \multicolumn{3}{|c}{Hausdorff Distance(mm) $\downarrow$} \\
        & Name & Details & Repeat & Size \\
        \hline
        & Input & - & - & $4\times128\times128\times 128$ \\
        \hline
        \multirow{8}{*}{Encoder} & InitConv & Conv3, Dropout & 1 & $16\times128\times128\times128$ \\
        \cline{2-5}
        & EnBlock1 & BN, ReLU, Conv3, BN, ReLU, Conv3, + & 1 & $16\times128\times128\times128$ \\
        \cline{2-5}
        & EnDown1 & Conv3 (stride 2) & 1 & $32\times64\times64\times64$ \\
        \cline{2-5}
        & EnBlock2 & BN, ReLU, Conv3, BN, ReLU, Conv3, + & 2 & $32\times64\times64\times64$\\
        \cline{2-5}
        & EnDown2 & Conv3 (stride 2) & 1 & $64\times32\times32\times32$ \\
        \cline{2-5}
        & EnBlock3 & BN, ReLU, Conv3, BN, ReLU, Conv3, + & 2 & $64\times32\times32\times32$\\
        \cline{2-5}
        & EnDown3 & Conv3 (stride 2) & 1 & $128\times16\times16\times16$ \\
        \cline{2-5}
        & EnBlock4 & BN, ReLU, Conv3, BN, ReLU, Conv3, + & 4 & $128\times16\times16\times16$\\
        \hline
        
        \multirow{8}{*}{Decoder} & DeUp3 & Conv3, ConvT, EnBlock3(Concate), Conv3& 1 & $64\times32\times32\times32$ \\
        \cline{2-5}
        & DeBlock3 & BN, ReLU, Conv3, BN, ReLU, Conv3, + & 1 & $16\times128\times128\times128$ \\
        \cline{2-5}
        & DeUp2 & Conv3, ConvT, EnBlock2(Concate), Conv3 & 1 & $32\times64\times64\times64$ \\
        \cline{2-5}
        & DeBlock2 & BN, ReLU, Conv3, BN, ReLU, Conv3, + & 1 & $32\times64\times64\times64$\\
        \cline{2-5}
        & DeUp1 & Conv3, ConvT, EnBlock2(Concate), Conv3 & 1 & $16\times128\times128\times128$ \\
        \cline{2-5}
        & DeBlock1 & BN, ReLU, Conv3, BN, ReLU, Conv3, + & 1 & $16\times128\times128\times128$\\
        \cline{2-5}
        & EndConv & Conv1 & 1 & $4\times128\times128\times128$ \\
        \cline{2-5}
        & Softmax & Softmax & 1 & $4\times128\times128\times128$ \\
        \hline
        
        \multirow{3}{*}{\makecell[c]{Attention \\ Conv Path}} & AttConv1 & Conv3 (stride 2) & 1 & $16\times64\times64\times64$ \\
        \cline{2-5}
        & AttConv2 & Conv3 (stride 2) & 1 & $16\times32\times32\times32$ \\
        \cline{2-5}
        & AttConv3 & Conv3 (stride 2) & 1 & $4\times16\times16\times16$ \\
        \hline
        
    \end{tabular}
    }
    
\end{table*}

\section{Experiments and Results Analysis}

\subsection{Datasets and Evaluation Metrics}
The dataset we used in the experiment is the 3D MRI data provided by the Brain Tumor Segmentation (BraTS) 2019 challenge \cite{menze2014multimodal, bakas2017advancing, bakas2018identifying, bakas2017segmentation}. It consists of four different MR sequences, namely native T1-weighted (T1), post-contrast T1-weighted (T1ce), T2-weighted (T2) and Fluid Attenuated Inversion Recovery (FLAIR). The volume of each sequence is $240\times240\times155$, and has been aligned in the same space. The labels used for tumor segmentation include background (label 0), necrotic and non-enhancing tumors (label 1), peritumoral edema (label 2), and GD-enhancing tumors (label 4). The dataset contains 335 patients for training and 125 for validation. Although the current test set result submission channel has been closed, the performance on validation set can still be verified by the online evaluation server.

Formally, the effectiveness is evaluated by the computational complexity and the segmentation accuracy. The complexity is determined by the amount of network parameters and FLOPs (i.e. multiplication and addition). The segmentation accuracy in our experiments is measured by the Dice score and the Hausdorff distance (95\%) for validation set. Moreover, We adopt Hausdorff distance (100\%) in the cross-validation experiment to show the impact of possible outliers. Each metric can be computed for the enhancing tumor region (ET, i.e. label 1), the whole tumor region (WT, i.e. labels 1, 2 and 4), and the regions of the tumor core (TC, i.e. labels 1 and 4).

\subsection{Implementation Details}
In our experiments, we use a batch size of 8 and train the CGA U-Net on 4 Nvidia Titan RTX GPUs (each has 24GB memory) for 500 epochs. The Adam optimizer is used with an initial learning rate of 0.001. We apply the following data augmentation techniques: (1) random cropping the MRI data from $240\times240\times155$ voxels to $128\times128\times128$ voxels; (2) random mirror flipping across the axial, coronal and sagittal planes by a probability of 0.5; (3) random intensity shift between $[-0.1, 0.1]$ and scale between $[0.9, 1.1]$. The softmax Dice loss is employed to train the network. $L2$ norm is applied for model regularization with a weight decay rate of $10^{-5}$. All of our experiments use the same initialization method and training from scratch. In addition, we fixed the seeds in all experiments to ensure that the results can be reproduced.

\subsection{Experimental Results}

We compare our method with the state-of-the-art approaches on the BraTS 2019 validation set. 
To be fair, we compare with the 3D methods and conduct experiments under the same parameter configuration. From Table.~\ref{tab:comparison}, our proposed GCA U-Net achieves dice scores of $78.83\%$, $89.29\%$, $82.32\%$ on ET, WT, TC, respectively. In particular, considerable gains ($1.73\%$ and $1.02\%$) have been achieved for the important ET and TC. Large improvements are also observed in terms of Hausdorff distance measure. Compared with Attention U-Net, our method shows great superiority in all metrics by supervising the attention map. Compared with nnU-Net, our method has a clear advantage on ET and slightly lower on TC and WT. However, the parameters and FLOPs of nnU-Net are 31.2M and 475.7G, respectively, while the corresponding values of our CGA U-Net are 4.77M and 151.12G. We obtain comparable results to nnU-Net with minimal computational complexity.

\begin{table}[!t]
\scriptsize
    \centering
    \caption{Performance comparison on the BraTS 2019 validation set.}
    \label{tab:comparison}
    \setlength{\tabcolsep}{3mm}{
    \begin{tabular}{l|c|c|c|c|c|c}

        \hline
        \multirow{2}{*}{Method} & \multicolumn{3}{c}{Dice Score (\%) $\uparrow$} & \multicolumn{3}{|c}{Hausdorff Distance(mm) $\downarrow$} \\
        \cline{2-7}
        & \bfseries ET & \bfseries WT & \bfseries TC & \bfseries ET & \bfseries WT & \bfseries TC\\
        \hline
        3D U-Net            & 70.86 & 87.38 & 72.48 & 5.062 & 9.432 & 8.719\\
        V-Net               & 73.89 & 88.73 & 76.56 & 6.131 & 6.256 & 8.705\\
        KiU-Net \cite{valanarasu2020kiu}             & 73.21 & 87.60 & 73.92 & 6.323 & 8.942 & 9.893\\
        Attention U-Net     & 75.96 & 88.81 & 77.20 & 5.202 & 7.756 & 8.258\\
        Wang et al. \cite{wang20193d}         & 73.70 & 89.40 & 80.70 & 5.994 & 5.677 & 7.357\\
        Li et al. \cite{li2019multi}           & 77.10 & 88.60 & 81.30 & 6.033 & 6.232 & 7.409\\
        nnU-Net \cite{isensee2021nnu}         & 74.80 & 90.81 & 83.90 & 3.77 & 4.36 & 6.42\\
        \bf{Ours}                & 78.83 & 89.29 & 82.32 & 3.262 & 5.809 & 6.721\\
        \hline
    \end{tabular}
    }
    
\end{table}

The parameter size and computational cost (FLOPs) of three different attention modules are compared in Table.~\ref{tab:complexity}. FLOPs generated by convolution and matrix multiplication are both taken into account. The parameters of the convolutional layer is $C_{in} \times K_h \times K_w \times K_d \times C_{out}$, where $K$ indicates the kernel size and $C$ refers to the number of channels. FLOPs generated by a convolution layer is calculated as $C_{in} \times K_h \times K_w \times K_d \times H \times W \times D \times C_{out}$, where $H, W, D$ are the size of the input feature map. Assuming that the dimensions of the two matrices are $A \times B$ and $B \times C$, then the FLOPs generated by the multiplication of these two matrices are $A \times B \times C$. The proposed SAM only has 0.675G FLOPs and 0.095M parameters, which is a very lightweight part of the whole network. Experiments show that our SAM module achieves better results with minimal computational overhead.

\begin{table}[ht]
\scriptsize
    \centering
    \caption{Computational complexity comparison of different attention modules.}
    \label{tab:complexity}
    
    \setlength{\tabcolsep}{5mm}{
    \begin{tabular}{l|c|c|c}
    \hline
        \bfseries Module & \bfseries Downsampling Rate & \bfseries FLOPs(G) & \bfseries Para.(M)\\
        \hline
        Self-Attention   &  8  & \makecell[c]{4.495} & \makecell[c]{0.492}\\
        CP Block         &  8  & \makecell[c]{2.284} & \makecell[c]{0.328}\\
        SAM(Ours)        &  8  & \makecell[c]{0.675} & \makecell[c]{0.095}\\
        SAM(Ours)        &  4  & \makecell[c]{1.97} & \makecell[c]{0.034}\\
        SAM(Ours)        &  2  & \makecell[c]{2.89} & \makecell[c]{0.01}\\
        \hline
    \end{tabular}
    }
\end{table}

In order to illustrate the sensitivity of SAM to the down-sampling rate, we calculate and analyse the parameters and FLOPs of SAM in Table.~\ref{tab:complexity}. The parameters of SAM mainly come from the convolution kernels of the attention convolution path. When the down-sampling rate of Ground Truth is reduced, the number of convolution kernels for down-sampling will be decreased, leading to a decline in the amount of SAM parameters. In order to ensure that we can learn an effective attention map, the number of convolution layers and channels can be adjusted flexibly. Therefore, the change of the parameters should be determined according to the actual situation. The main growth of FLOPs comes from the convolution part with the increase of the size of Ground Truth, but it will not boost as sharply as matrix multiplication.

\begin{figure*}[ht]
 % Caption and label go in the first argument and the figure contents
 % go in the second argument
    \centering
     % \vspace{-0.5cm}
     % \setlength\belowdisplayskip{3pt}
    \includegraphics[width=1\textwidth]{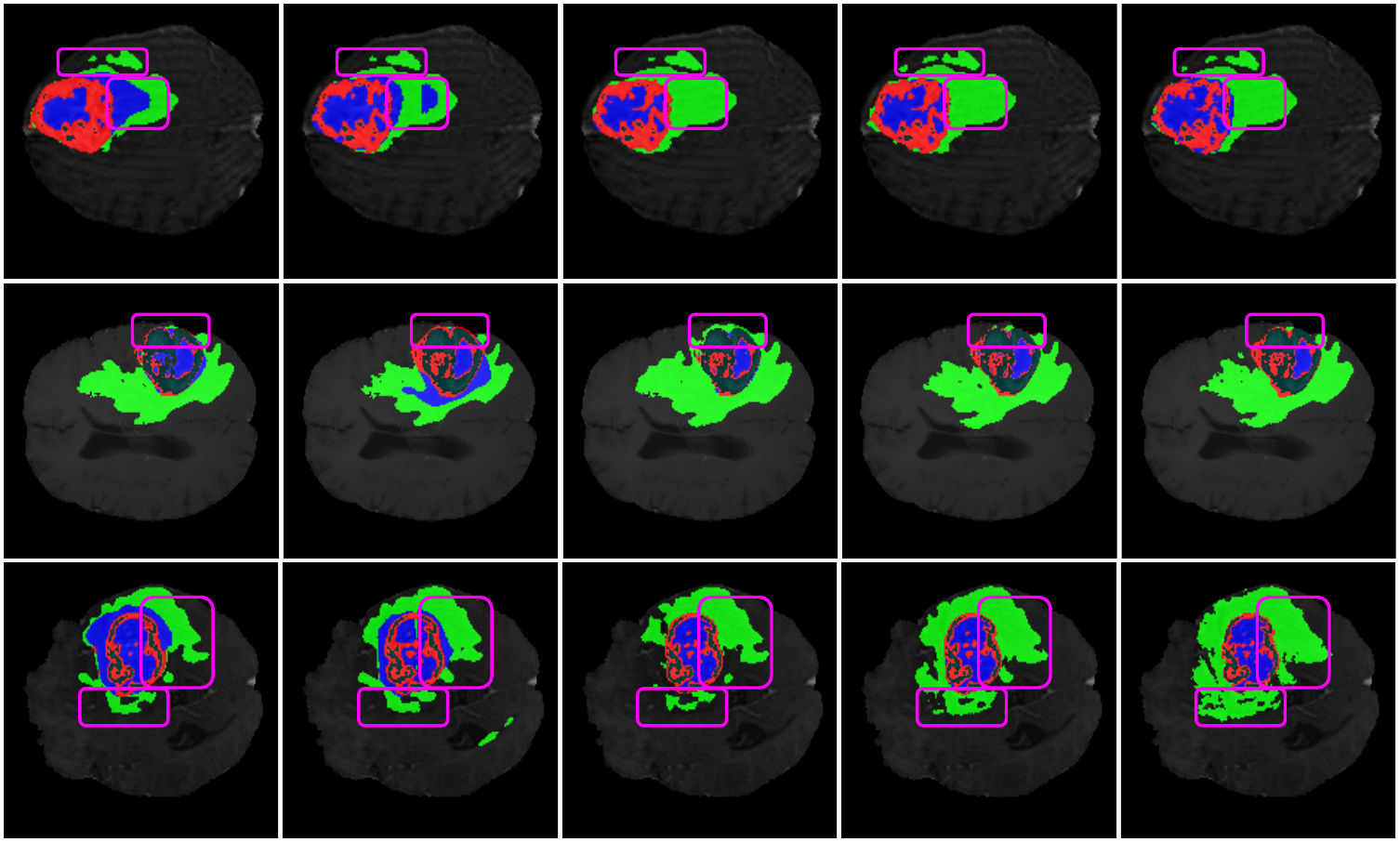}
    \begin{tabu} to 0.85\linewidth{ X[1.0c] X[1.0c] X[1.0c] X[1.0c] X[1.0c]}
        U-Net3D & VNet & \small{Attention U-Net} & Ours & Ground Truth \\
    \end{tabu}
    \caption{The visual comparison of MRI brain tumor segmentation results.}
    \label{fig:vis-prediction}
\end{figure*}

In addition, we provide visualization of segmentation results of different methods in Figure.~\ref{fig:vis-prediction}. It should be noted that in the visualization experiment, we select 80\% of the data for training, and the remaining 20\% for testing and visualization.  Since our method can capture stable and accurate global context information, it is very effective in repairing missing tumor details and suppressing over-segmentation.

\subsection{Ablation Study}

\begin{figure*}[ht]
    \centering
    \includegraphics[width=0.65\textwidth]{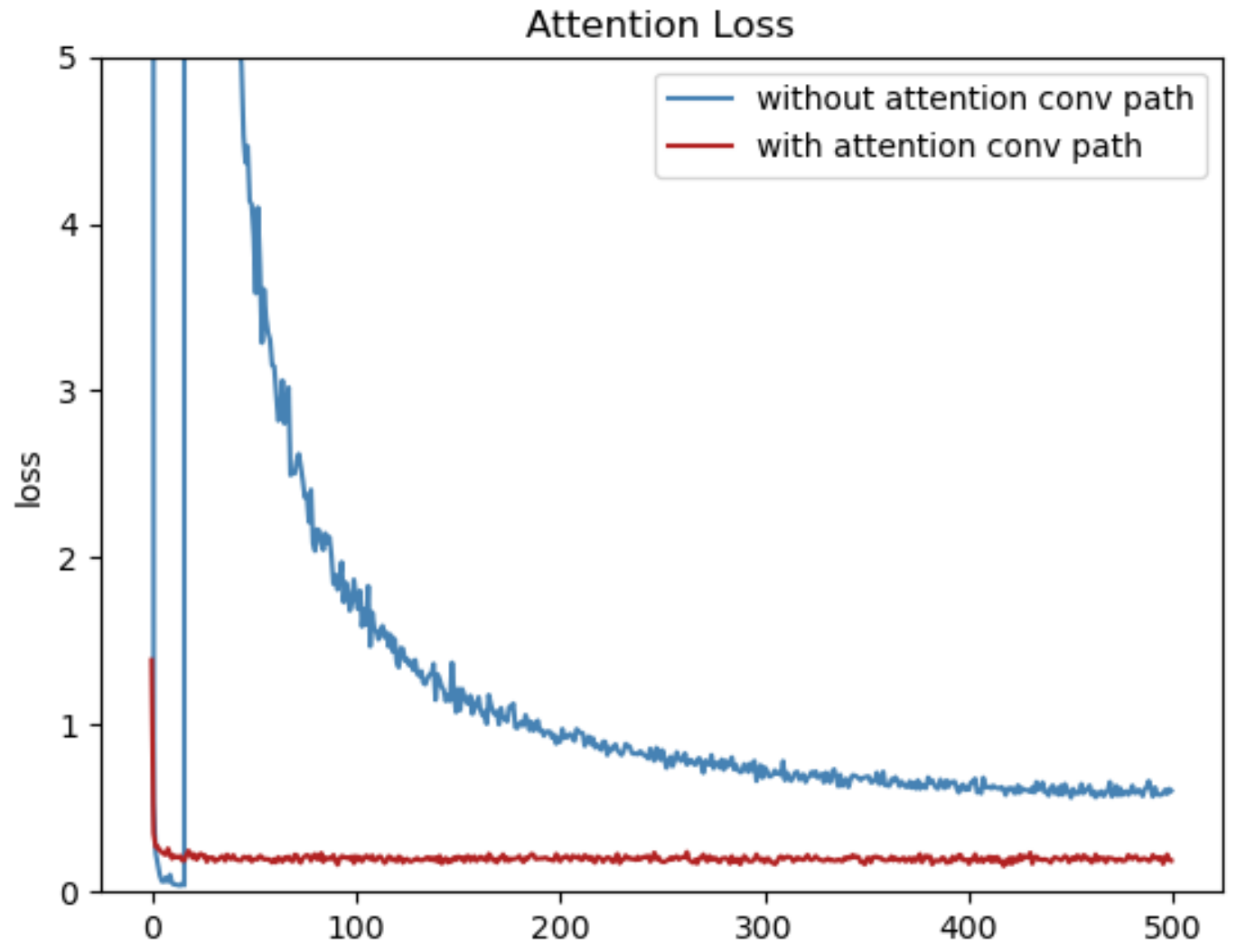}
    \caption{The visual comparison of the attention loss during training with/without attention conv path.}
    \label{fig:attention-loss}
\end{figure*}

\textbf{Attention Convolution Path.}
In order to compare the way the CP Layer obtains the attention map and our Attention Convolution Path, we record the attention map loss during the training process. At first, we remove the attention conv path and use a convolution layer (which CP Block used) to obtain the attention map. As shown by the blue line in Figure.~\ref{fig:attention-loss}, the change of the attention loss without attention conv path during the training process was plotted. In addition, we let the red line represent the attention loss curve using our attention conv path. It can be clearly seen that the blue line fluctuates significantly in the early stages of training and finally converges to a larger value. Our attention loss can quickly converge to the ideal state during training. This result proves that using the CP map way to obtain the attention map may have optimization problems during training process. Actually, the Main Loss is difficult to converge without the attention conv path. At the same time, the structure of the attention conv path can also be flexibly adjusted to adapt to different situations.

\begin{table*}[ht]
    \scriptsize
    \centering
    \setlength{\tabcolsep}{3mm}{
        \begin{tabular}{l|c|c|c|c|c|c}
        \hline
        \multirow{2}{*}{Method} & \multicolumn{3}{c}{Dice Score (\%) $\uparrow$} & \multicolumn{3}{|c}{100\% Hausdorff Distance(mm) $\downarrow$} \\
        \cline{2-7}
         & \makecell[c]{\bfseries ET\\ class4} & \makecell[c]{\bfseries WT\\ class1\&2\&4} & \makecell[c]{\bfseries TC\\ class1\&2} & \makecell[c]{\bfseries ET\\ class4} & 
        \makecell[c]{\bfseries WT\\ class1\&2\&4} & \makecell[c]{\bfseries TC\\ class1\&2}\\
        \hline
        Baseline                    & 74.47 & 88.35 & 79.75 & 29.860 & 52.748 & 28.481\\
        Baseline + Intra            & 78.60 & \textbf{90.07} & 81.66 & 18.173 & 35.034 & 25.300\\
        Baseline + Intra (class1)   & 77.16 & 89.92 & 81.17 & 23.448 & 40.934 & 24.253\\
        Baseline + Intra (class2)   & 77.64 & 89.03 & 82.30 & 22.962 & 50.669 & 24.042\\
        Baseline + Intra (class4)   & 77.96 & 89.56 & 80.91 & 24.021 & 46.688 & 27.859\\
        Baseline + Inter            & 78.09 & 89.33 & 82.54 & 18.783 & 45.704 & 19.938\\
        Baseline $-$ Inter          & 73.71 & 87.57 & 78.21 & 35.674 & 53.793 & 29.140\\
        Baseline + Intra + Inter    & \textbf{79.21} & 89.8 & \textbf{83.46} & \textbf{17.418} & \textbf{32.440} & \textbf{19.833}\\
        \hline
        \end{tabular}
    }
    \caption{Results of the 5-fold cross-validation on the BraTS 2019 training set.}
    \label{tab:ablation}
\end{table*}

\textbf{Intra-class and Inter-class methods.}
In order to verify the effectiveness of intra-class update method and inter-class distance optimization, we conduct ablation experiments as illustrated in Table.~\ref{tab:ablation}. We make some modifications to the U-Net and use it as the baseline in our experiments. The details of the network can be found in Table.~\ref{tab:network_details}. Baseline + Intra indicates using the intra-class update method to reconstruct all the classes in feature maps. For a more detailed category update effect, we conduct three extra experiments by updating only one class in the feature map (e.g. class1, class2 or class4) with the help of intra-class update. The performance comparison with the intra-class update method for different classes demonstrates that the SAM can improve the result for each corresponding class. For example, the experiment only updating class4 has the most obvious improvement on the accuracy for ET. Moreover, the inter-class distance optimization method also boosts the Dice score for ET and TC. Using the two methods together can achieve the best result which has an improvement of $4.74\%$, $1.45\%$, $3.71\%$ for ET, WT and TC, respectively.

Moreover, in order to analyze whether our inter-class distance optimization can really increase the distance between different categories, we reverse the sign of Equation 10 during the training process and performed an Baseline $-$ Inter experiment as shown in the second last row of Table.~\ref{tab:ablation}. Therefore, the optimization direction of $L_I$ is to minimize the inter-class distance. Since the $L_I$ is not used for back propagation in the first 20 epochs, the inter-class distance will gradually increase as the model learns. After 20 epochs, $L_I$ join the back propagation and make the inter-class distance decrease and converge. Under the influence of this reverse optimization, the prediction results deteriorate significantly.

\begin{figure}[htb]
 % Caption and label go in the first argument and the figure contents
 % go in the second argument
    \centering
    % \vspace{-0.5cm}
    % \setlength\belowdisplayskip{3pt}
    \includegraphics[width=0.9\linewidth]{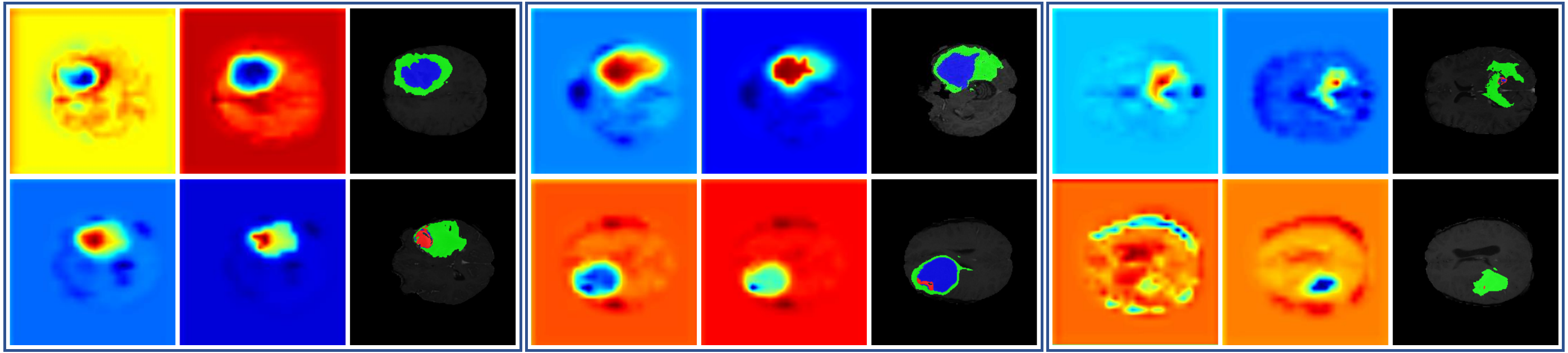}
    \begin{tabu} to 0.9\linewidth{ X[1.0c] X[1.0c] X[1.0c] X[1.0c] X[1.0c] X[1.0c] X[1.0c] X[1.0c] X[1.0c]}
     \small{Before} & \small{After} & \small{GT} & \small{Before} & \small{After} & \small{GT} & \small{Before} & \small{After} & \small{GT}
    \end{tabu}
    \caption{Visualization of feature map before and after update by SAM. GT: Ground Truth. Redder point indicates a larger value, and bluer point indicates a smaller value. }
    \label{fig:vis-feature}
\end{figure}

To further explore the role of SAM in reconstructing feature maps, we visualize the feature maps before and after reconstruction according to channels using both two methods in Figure.~\ref{fig:vis-feature}. These visualizations can be regarded as heatmaps, which red indicates a larger value and blue indicates a smaller value. Compared with the ‘before’ image, for all pixels belonging to the background, the value in ‘after’ image has a higher degree of similarity and confidence in this category. As a result, there will be a clearer outline in the visualization and it is also closer to the shape of the label. 

Moreover, the tumor regions in the feature maps have different values compared to other parts and display in different colors after visualization. We can see that the updated feature map has a clearer outline and inter-class contrast in the tumor region. In addition, the contrast between other areas of the brain (background) is decreased, which significantly reduces interference with tumor segmentation.

\subsection{Attention Map \& Prediction Visualization}

\begin{figure}[htb]
 % Caption and label go in the first argument and the figure contents
 % go in the second argument
 \vspace{-\intextsep}
    \centering
      \vspace{0.5cm}
    \includegraphics[width=0.8\linewidth]{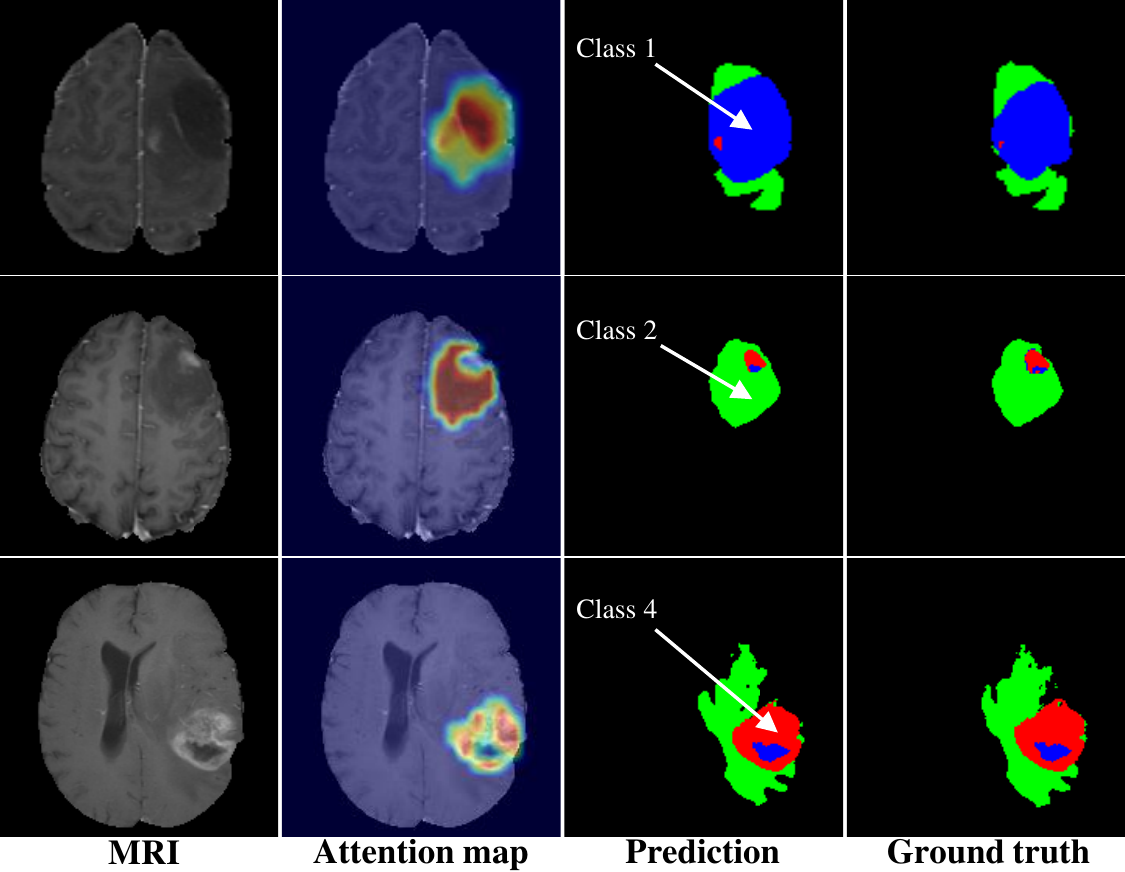}
    \caption{Visualization of the supervised attention map. The three rows visualize the category distribution map corresponding to class 1, 2 and 4, respectively.
    }
    \label{fig:vis-heatmap}
    %\vspace{-20pt}
\end{figure}

To verify that making the attention map under supervision has positive effects, we show a visual comparison of attention maps for different classes in Figure.~\ref{fig:vis-heatmap}. The visualization of the attention map clearly shows the position of the corresponding category, which greatly facilitates the segmentation. Even small areas like non-enhancing tumors can be captured by the SAM. These experiments validate that the supervision of the map can effectively encode stable global context information for guiding the network to predict accurate results.

\subsection{Method Robustness}
The reliability of deep learning systems for medical images depends on their accuracy but also on their robustness. A method that is highly accurate but fails dramatically in a few cases may lack of trust from medical experts. Therefore, we have counted the standard deviations of different methods in ablation study to compare their differences in stability as shown in Table.~\ref{tab:stddev}.

\begin{table*}[ht]
\scriptsize
    \centering
    \caption{Standard deviation results of the 5-fold cross-validation on the BraTS 2019 training set.}
    \label{tab:stddev}
    \setlength{\tabcolsep}{3mm}{
    \begin{tabular}{l|c|c|c|c|c|c}
        \hline
        \multirow{2}{*}{Method} & \multicolumn{3}{c}{StdDev of Dice Score (\%) $\downarrow$} & \multicolumn{3}{|c}{StdDev of 100\% Hausdorff Distance(mm) $\downarrow$} \\
        \cline{2-7}
         & \makecell[c]{\bfseries ET\\ class4} & \makecell[c]{\bfseries WT\\ class1\&2\&4} & \makecell[c]{\bfseries TC\\ class1\&2} & \makecell[c]{\bfseries ET\\ class4} & 
         \makecell[c]{\bfseries WT\\ class1\&2\&4} & \makecell[c]{\bfseries TC\\ class1\&2}\\
        \hline
        Baseline                    & 30.28 & 7.47 & 13.68 & 23.38 & 25.91 & 25.48\\
        Baseline + Intra            & 29.14 & 6.71 & \textbf{10.41} & 21.60 & 22.19 & 20.16\\
        Baseline + Intra (class1)   & 26.75 & 6.57 & 13.56 & 17.56 & 24.33 & 17.80\\
        Baseline + Intra (class2)   & 28.21 & 6.80 & 14.87 & 21.49 & 25.24 & 23.69\\
        Baseline + Intra (class4)   & 28.17 & \textbf{6.11} & 12.70 & 16.42 & 26.58 & \textbf{17.75}\\
        Baseline + Inter            & 29.25 & 7.23 & 12.81 & 20.17 & 25.97 & 22.26\\
        Baseline + Intra + Inter    & \textbf{27.71} & 6.43 & 12.11 & \textbf{16.15} & \textbf{21.49} & 20.23\\
        \hline
    \end{tabular}
    }
\end{table*}

Since our method suppresses over-segmentation and under-segmentation to a certain extent, some extreme cases with low ET scores are avoided. Therefore, our method has a better performance in standard deviation, which makes the system more stable in the face of real complicated situation.

In addition, we further analyzed the Hausdorff distance metric in the ablation study. The Hausdorff distance metric set by the challenge organizer is 95\%, which is intended to eliminate the influence of a subset of outliers. Actually, the 5\% outliers may play an important role in creating radiation therapy planing considering their size and location. Therefore, We calculated the 100\% Hausdorff distance in the cross-validation experiment as illustrated in Table.~\ref{tab:ablation}. Due to outliers, the Hausdorff distance of 100\% is numerically larger than that of 95\%. But compared to the baseline, our method has played a more obvious role in these three categories. We can see that the Hausdorff distance of ET has dramatically decreased with the help of both the two methods.

\begin{figure}[htb]
    \centering
    % \vspace{-0.5cm}
    % \setlength\belowdisplayskip{3pt}
    \includegraphics[width=\linewidth]{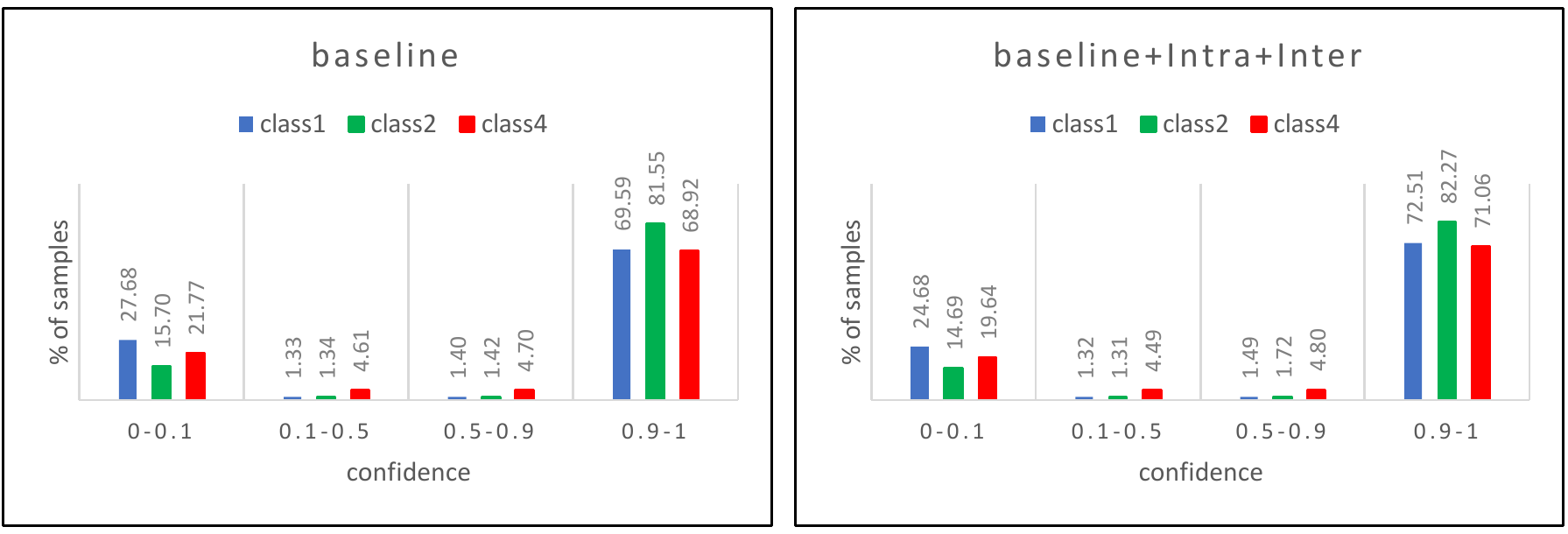}
    \caption{confidence distribution of model predictions of the 5-fold cross-validation on the BraTS 2019 training set.}
    \label{fig:confidence}
\end{figure}

The confidence of model prediction is also an important aspect of evaluating its robustness. We counted the confidence distribution of model predictions in the cross-validation experiment. Specifically, we divide the predicted probability into 10 intervals, and plot the predicted proportion of each interval in Fig \ref{fig:confidence}. Since the probability is mainly distributed in the two intervals of $[0, 0.1]$ and $[0.9, 1.0]$, we merge the middle interval into $[0.1, 0.5]$ and $[0.5, 0.9]$. Compared with the baseline, our method has more prediction probabilities that fall within the interval of $[0.9, 1.0]$, which means a higher degree of confidence. In addition, the decrease in the proportion of the $[0, 0.5]$ interval also indicates that the model discards those vacillating predictions.

\section{Conclusion}
We have proposed a novel framework named CGA U-Net to automatically segment brain tumor from MRI. The SAM significantly improved the segmentation results with only 0.095M parameters and around 0.675G FLOPs. To reconstruct the feature map accurately and efficiently, we designed intra-class update method and inter-class distance optimization. The experimental results on BraTS 2019 dataset show that our approach is highly effective. The final dice scores (78.83\%, 89.29\%, 82.32\% for ET, WT and TC, respectively) are superior or comparable to that of the state-of-the-art methods. %Moreover, we implemented our method on LiTS 2017 CT dataset and also achieved satisfactory results, which proved the generalization of it.

\section{Future Scope}
There may be some possible limitations in this study. Since we update the feature map based on the category, the computational complexity may expand as the number of categories increases. This work may consider dataset with larger number of categories and conduct more detailed intra-class update experiments. The future work can also include sensitivity analysis on the number of categories. In addition, this work may consider more datasets such as the decathlon challenge to verify the generalization of the method.

\section{Reference}

\bibliographystyle{iopart-num}
\bibliography{references}

\end{document}